\documentclass[aps,prc,reprint,superscriptaddress]{revtex4-1}
\usepackage{dcolumn}
\usepackage{amsmath,amssymb,amsthm,paralist}
\usepackage{graphicx}
\usepackage{appendix}
\usepackage{amsmath}
\usepackage{amsfonts}
\usepackage{hyperref}
\usepackage{url}
\usepackage{verbatim}

\usepackage{floatrow}
\UseRawInputEncoding 

\begin{document}
\title{Determination of energy-dependent neutron backgrounds using shadow bars}

\author{S. N. Paneru}
\email[]{paneru@frib.msu.edu}

\affiliation{Facility for Rare Isotope Beams, East Lansing, Michigan 48824, USA}
\author{K. W. Brown}
\affiliation{Facility for Rare Isotope Beams, East Lansing, Michigan 48824, USA}
\affiliation{Department of Chemistry, Michigan State University, East Lansing, MI 48824, USA}
\author{F.C.E Teh}
\affiliation{Facility for Rare Isotope Beams, East Lansing, Michigan 48824, USA}
\affiliation{Department of Physics and Astronomy, Michigan State University, East Lansing, MI 48824, USA}
\author{K. Zhu}
\affiliation{National Superconducting Cyclotron Laboratory, Michigan State University, East Lansing, Michigan 48824, USA}
\affiliation{Department of Physics and Astronomy, Michigan State University, East Lansing, MI 48824, USA}
\author{M.B. Tsang}
\affiliation{Facility for Rare Isotope Beams, East Lansing, Michigan 48824, USA}
\affiliation{Department of Physics and Astronomy, Michigan State University, East Lansing, MI 48824, USA}
\author{D. Dell'Aquila}
\affiliation{National Superconducting Cyclotron Laboratory, Michigan State University, East Lansing, Michigan 48824, USA}
\author{Z. Chajecki}
\affiliation{Department of Physics, Western
Michigan University, Kalamazoo, MI 49008, USA}
\author{W.G. Lynch}
\affiliation{Facility for Rare Isotope Beams, East Lansing, Michigan 48824, USA}
\affiliation{Department of Physics and Astronomy, Michigan State University, East Lansing, MI 48824, USA}
\author{S. Sweany}
\affiliation{National Superconducting Cyclotron Laboratory, Michigan State University, East Lansing, Michigan 48824, USA}
\affiliation{Department of Physics and Astronomy, Michigan State University, East Lansing, MI 48824, USA}
\author{C.Y. Tsang}
\affiliation{National Superconducting Cyclotron Laboratory, Michigan State University, East Lansing, Michigan 48824, USA}
\affiliation{Department of Physics and Astronomy, Michigan State University, East Lansing, MI 48824, USA}
\author{A.K. Anthony}
\affiliation{Facility for Rare Isotope Beams, East Lansing, Michigan 48824, USA}
\affiliation{Department of Physics and Astronomy, Michigan State University, East Lansing, MI 48824, USA}
\author{J. Barney}
\affiliation{National Superconducting Cyclotron Laboratory, Michigan State University, East Lansing, Michigan 48824, USA}
\affiliation{Department of Physics and Astronomy, Michigan State University, East Lansing, MI 48824, USA}
\author{J. Estee}
\affiliation{National Superconducting Cyclotron Laboratory, Michigan State University, East Lansing, Michigan 48824, USA}
\affiliation{Department of Physics and Astronomy, Michigan State University, East Lansing, MI 48824, USA}
\author{I. Gasparic}
\affiliation{Division of Experimental Physics, Rudjer Boskovic Institute, Zagreb, Croatia}
\author{G. Jhang}
\affiliation{National Superconducting Cyclotron Laboratory, Michigan State University, East Lansing, Michigan 48824, USA}
\author{O.B. Khanal}
\affiliation{Department of Physics, Western
Michigan University, Kalamazoo, MI 49008, USA}
\author{J. Mandredi}
\affiliation{National Superconducting Cyclotron Laboratory, Michigan State University, East Lansing, Michigan 48824, USA}
\affiliation{Department of Physics and Astronomy, Michigan State University, East Lansing, MI 48824, USA}
\author{C.Y. Niu}
\affiliation{National Superconducting Cyclotron Laboratory, Michigan State University, East Lansing, Michigan 48824, USA}
\author{R.S. Wang}
\affiliation{National Superconducting Cyclotron Laboratory, Michigan State University, East Lansing, Michigan 48824, USA}
\author{J.C. Zamora}
\affiliation{Facility for Rare Isotope Beams, East Lansing, Michigan 48824, USA}

\begin{abstract}
Understanding the neutron background is essential for determining the neutron yield from nuclear reactions. In the analysis presented here, the shadow bars are placed in front of neutron detectors to determine the energy dependent neutron background fractions. The measurement of neutron spectra with and without shadow bars is important to determine the neutron background more accurately. The neutron background, along with its sources and systematic uncertainties, are explored with a focus on the impact of background models and their dependence on neutron energy. 
\end{abstract}


\maketitle
\section{Introduction}
Neutron detection provides essential insights into the nuclear dynamics of a nuclear reaction and nuclear structure. As the availability of access to radioactive beams has increased, allowing access to more neutron rich nuclei, the problem of properly measuring the neutron yields has become increasingly important.Thus more dedicated neutron arrays~\cite{MONA,NEDA,VANDLE,NeuLAND,NEBULA} have become available in recent years. The problem of measuring neutron yields is complicated for many reasons such as neutron background, low neutron detection efficiency, varying behaviour of neutrons with energy, etc. A critical component in the analysis of neutron yields is the ability to accurately model and subtract the neutron background from the experimentally measured yield.

Usually, a bar or block, called a ``shadow bar" hereafter, is placed between the target and the neutron detector to prevent neutrons from taking a direct path to reach the detector. Comparing the number of neutrons detected in the blocked and unblocked areas provides an estimate of the neutron background. This method is limited by the assumption that the estimated background is independent of neutron energy and the neutron distribution is isotropic. The previous works on neutron background determination  in Refs.~\cite{VANDLE,LENDA,Heilbronn,Keita} do not explore the consequences of these assumptions. In this paper, we determine the energy dependent neutron background using the shadow bar for the Large Area Neutron Array (LANA)~\cite{LANAref} constructed at the National Superconducting Cyclotron Laboratory. This technique can be applied to other neutron detectors. 
\par

\section{EXPERIMENT}\label{experiment}
The Large Area Neutron Array (LANA) is composed of two large
neutron walls, each possessing 25 independent detection cells that cover a total area of 2 x 2 $\text{m}^2$. Each detector cell is a 76$''$-long Pyrex glass container, with a rectangular cross section of height $h$ = 7.62 cm and depth $w$= 6.35 cm, filled with liquid organic scintillator NE-213. To maximize detection efficiency, LANA features large active detection area, small dead space between scintillators and minimum amount of inactive mass through which the neutrons must pass. For heavy ion collision experiments, it is designed to detect neutron from about 10 MeV to about 300 MeV when it is placed at 4.5 m from the target. LANA has been used in many previous nuclear experiments, for example in Ref ~\cite{Coupland,Fam06}. The analysis reported here will focus on a single neutron wall.\par
\begin{figure}[htp]
\captionsetup{}
\includegraphics[width=1.0\columnwidth,angle=0]{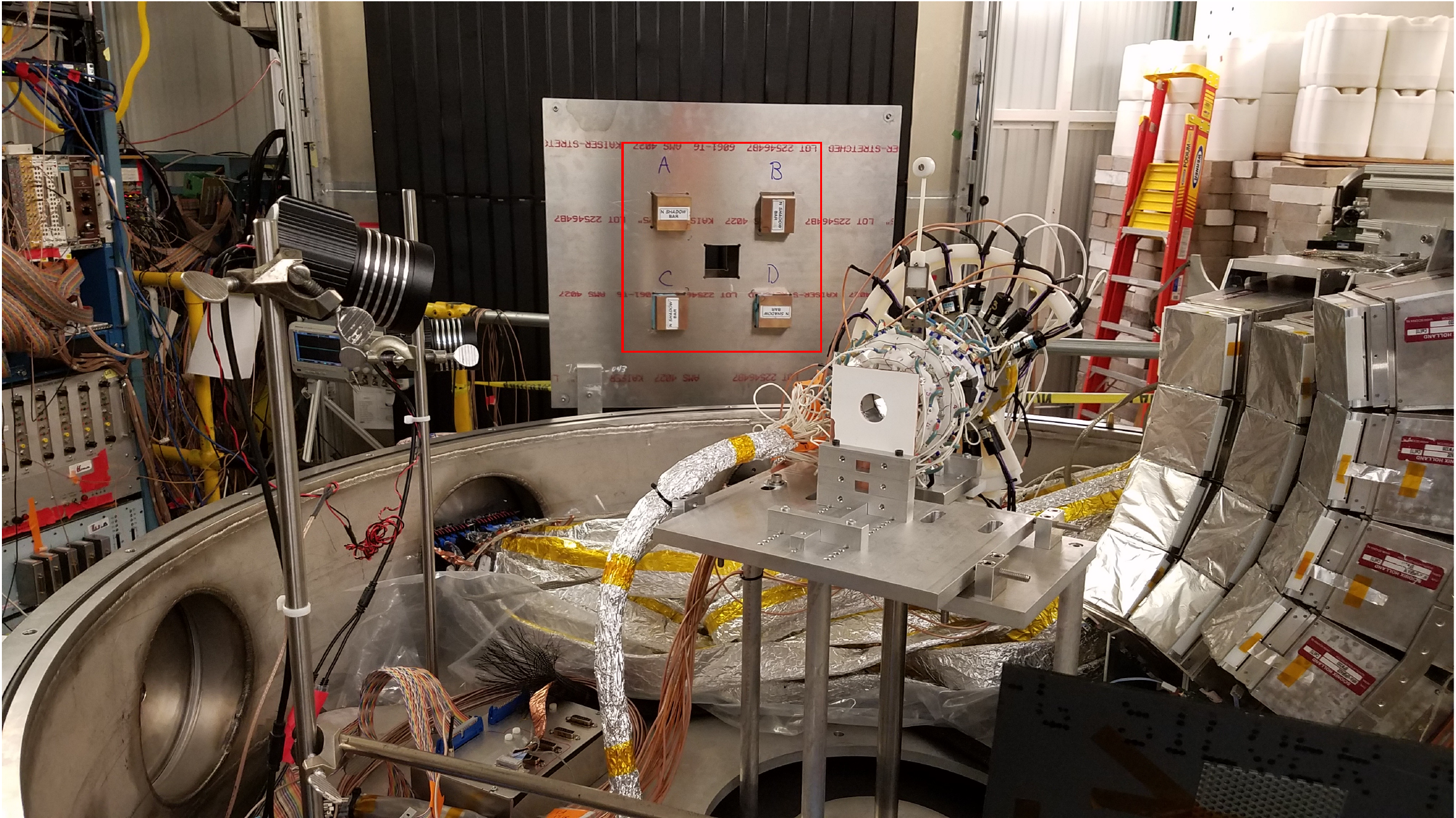}
\caption{Experimental Setup showing the shadow bar. The red rectangular box indicates the position of shadow bars.}
\label{setup}
\end{figure}
In the experiment, LANA was used to measure neutrons with beams of $^{40,48}$Ca at 56,140 MeV/u on targets of $^{58,64}$Ni and $^{112,124}$Sn. The calibration of LANA was performed using cosmic rays and an AmBe neutron source as explained in Ref.~\cite{zhu2020}. The neutrons were distinguished from charged particles using the charged particle veto wall consisting of thin bars of plastic scintillator that were placed in front of the LANA. The neutron-gamma separation was performed with value-assigned pulse shape discrimination method as explained in Ref.~\cite{fanurs}. The energy of neutrons is determined using the time of flight method.\par
\begin{figure}[htp]
\includegraphics[width=1.0\columnwidth,angle=0]{
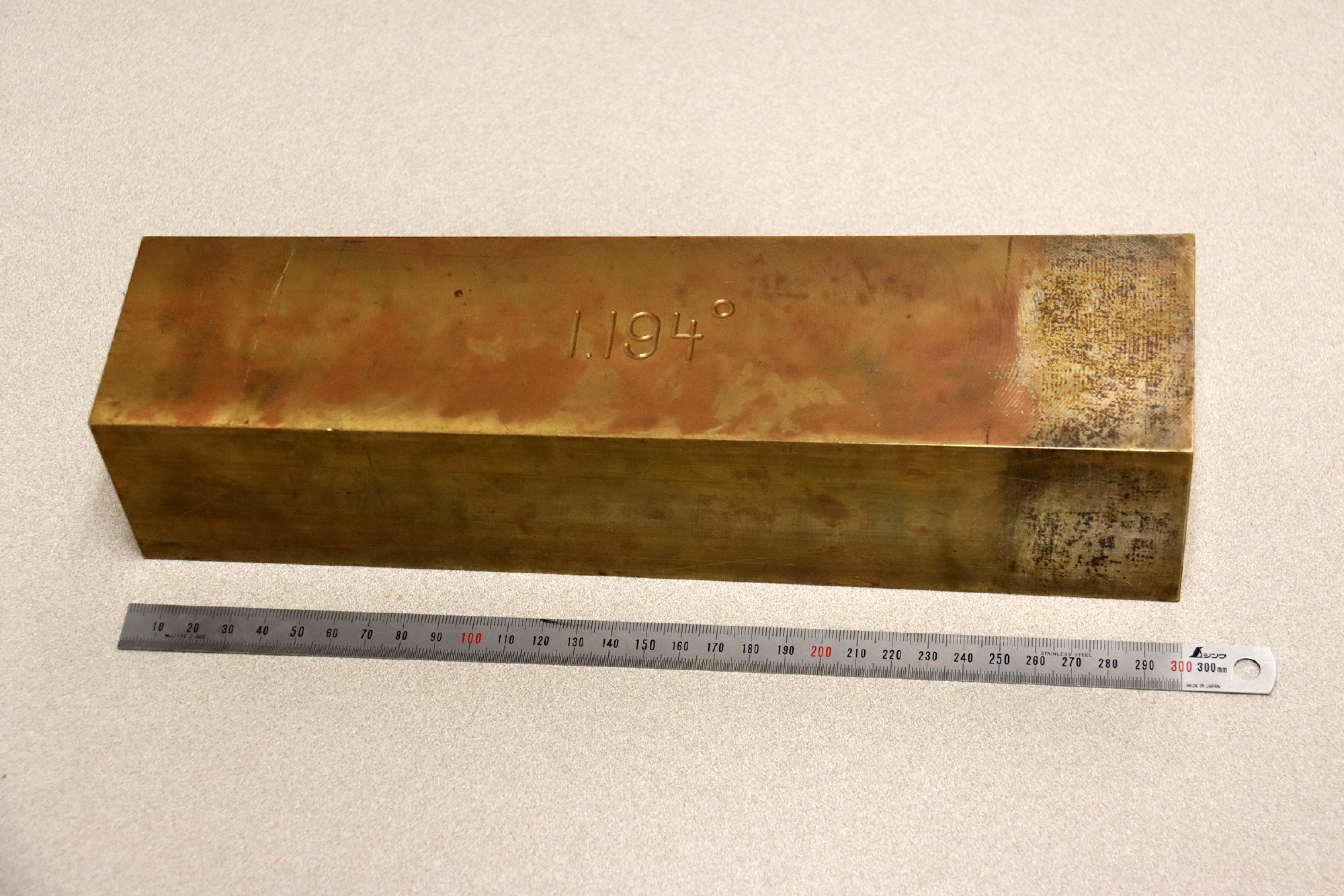}
\caption{Shadow bar used in experiment.}
\label{bar}
\end{figure}
\begin{figure}
\centering
\includegraphics[width=1.0\columnwidth,angle=0]{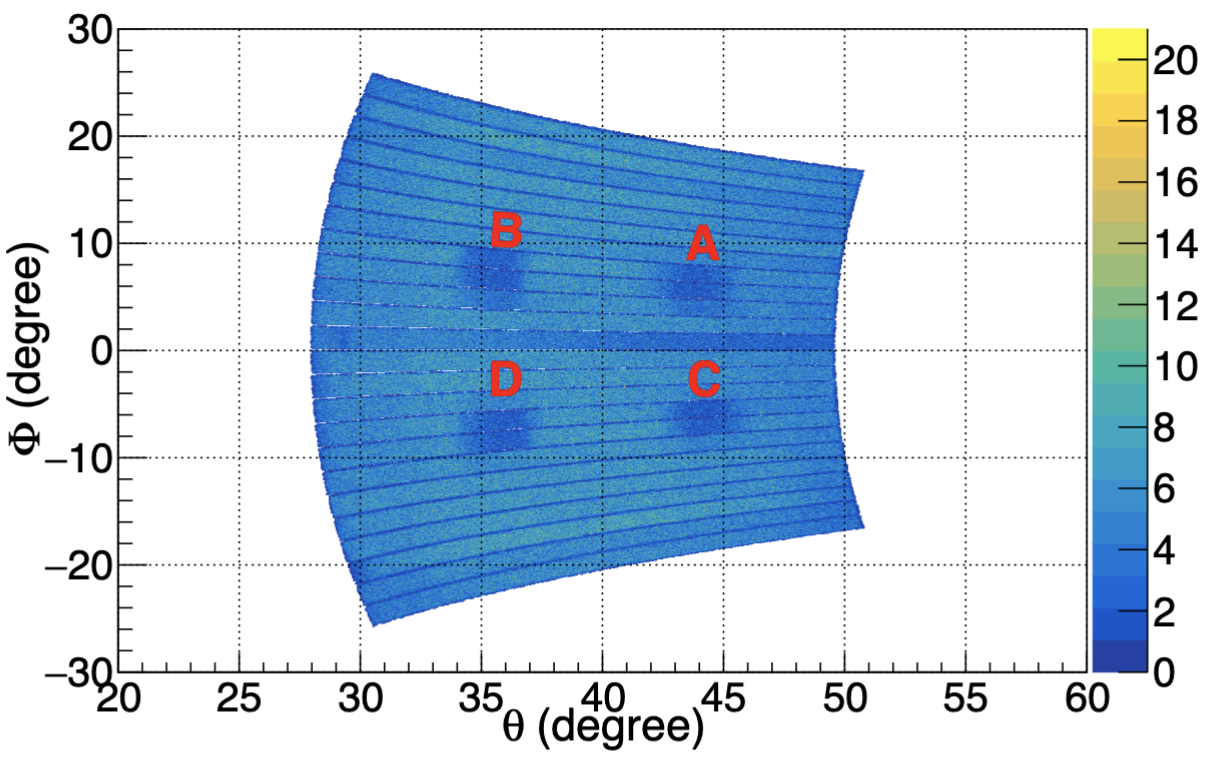}
\caption{The neutron hit pattern with 4 shadow bar installed in $\theta$-$\phi$ lab coordinates. The 4 square areas with less yield correspond to the area shadowed by the shadow bars. The neutron bars are counted from bottom to the top. The shadow bars at the bottom fully cover bar 8 and partially cover bars 7 and 9, while the shadow bars at the top fully cover bar 16 and partially cover bars 15 and 17. The reaction system for this plot is
$^{48}$Ca +$^{124}$ Sn at E/A = 140 MeV .}
\label{phi-theta}
\end{figure}

 To determine the neutron background, we block some areas of LANA and then compare the number of neutrons detected in blocked and not blocked areas of LANA. Four 30-cm-long brass shadow bars are used to block about 2$\%$ of front LANA detection area. They are placed about 1.7 m from the target, outside the chamber as shown in Fig.~\ref{setup}. The shadow bars are thick enough to stop neutrons of energies $\leq$300 MeV that take a direct path to the neutron wall. Figure~\ref{bar} shows one of the four shadow bars used in the experiment. The bottom bar of LANA is not used in our experiment because of the shadowing by stand holding the shadow bars.  Figure~\ref{phi-theta} shows the neutron hit distribution obtained for $^{48}$Ca beam impinging on $^{124}$ Sn at E/A = 140 MeV in lab $\theta$-$\phi$ coordinates on LANA, showing the decreased amount of neutrons in the shadow regions. The positions A, B, C, and D represent the shadow bar positions as shown in Fig.~\ref{setup}. The neutron bars are counted from bottom to the top. The shadow bars at positions C and D fully cover the vertical cross section of bar 8 and partially cover bars 7 and 9, while the shadow bars at positions A and B fully cover bar 16 and partially cover bars 15 and 17. The shadow bars in position A and C cover larger angles in $\theta$-$\phi$ distribution while the shadow bars in positions B and D cover more forward angles. The neutron spectrum is mostly unaffected by the thin aluminum mount holding the shadow bar as their interaction probability in this material is small and the energy loss of neutrons is also negligible. Example spectra for the detected neutron yield as a function of angle for one of the shadowed detector bars are shown in Fig.~\ref{neutron_spectra}, both with and without the shadow bars in place.

\begin{figure}[htp]
\captionsetup{}
\includegraphics[width=1.0\columnwidth,angle=0]{
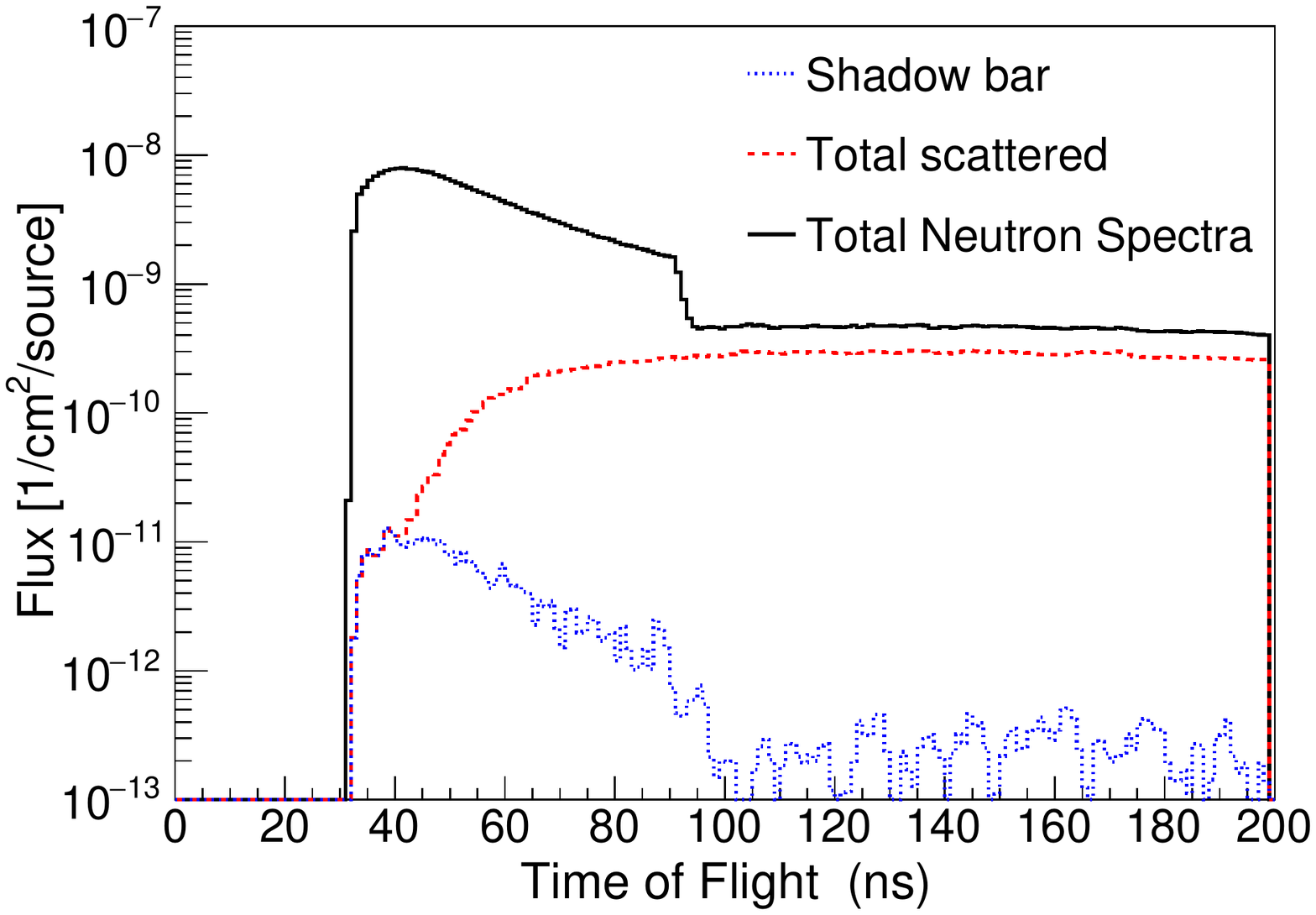}
\caption{The time of flight spectra deduced from the Monte Carlo simulation of neutron scattering. The blue dotted histogram shows the contribution of neutron scattering of the shadow bar. The red dashed histogram shows the total neutron scattering contribution from roof, walls and beam dump. The solid black histogram shows the neutron spectra observed from simulation.}
\label{monte_carlo}
\end{figure}

 Neutrons produced in the reaction target can also be scattered on the room walls (or other volume), and later detected in LANA. Any neutrons other than the ones directly hitting the LANA are considered as background neutrons. In order to investigate the various sources of neutron multiple scattering, a Monte-Carlo simulation was performed with the code PHITS \cite{phits}. A realistic geometry of the experimental hall, with the most representative scattering elements, was included in the simulation. Primary neutrons were generated in the target location by using the experimental energy distribution. The neutron angles were uniformly distributed in the 4$\pi$ solid angle. Secondary neutron  reactions were accounted in the simulation with the respective nucleon-nucleus models \cite{phim1,phim2}. Neutrons were tagged on event-by-event basis to separate the multiple-scattering component from the time of flight (ToF) spectra. The simulated ToF spectra for various neutron sources are shown in Fig.~\ref{monte_carlo}. The sharp edges in the total ToF spectrum (around 30 and 90 ns) correspond to the limits of the energy distribution used to generate the neutrons (between 20 to 200 MeV). 
The neutrons scattered off the shadow bars and hitting LANA could be contributing to the background. However, the simulation results show that the fraction of neutrons detected from this process contribute a very small component ($\sim$1/1000) to the total neutron spectrum (see dotted blue histogram in Fig.~\ref{monte_carlo}). Another major source of background is the scattering of  fast neutrons on the concrete volumes of the experimental hall such as floor, ceiling, walls, etc. The simulation estimated this type of component to only contribute at small neutron energies (or equivalently large time of flight) as shown by the dashed red histogram in Fig.~\ref{monte_carlo}. Neutrons produced in the beam dump have a longer ToF and do not contribute significantly to the neutron background, since they are located at the low energy region of the distribution.
The dominant background background contribution thus arise from secondary neutrons resulting from the charged particle induced reactions on various components of the experimental setup. The present simulation has a limitation to quantify realistically  this contribution given that the number of reaction channels that would produce neutrons is relatively large. \par

\section{Analysis}\label{algorithm}

\par
\begin{figure}[htp]
\includegraphics[width=1.0\columnwidth,angle=0]{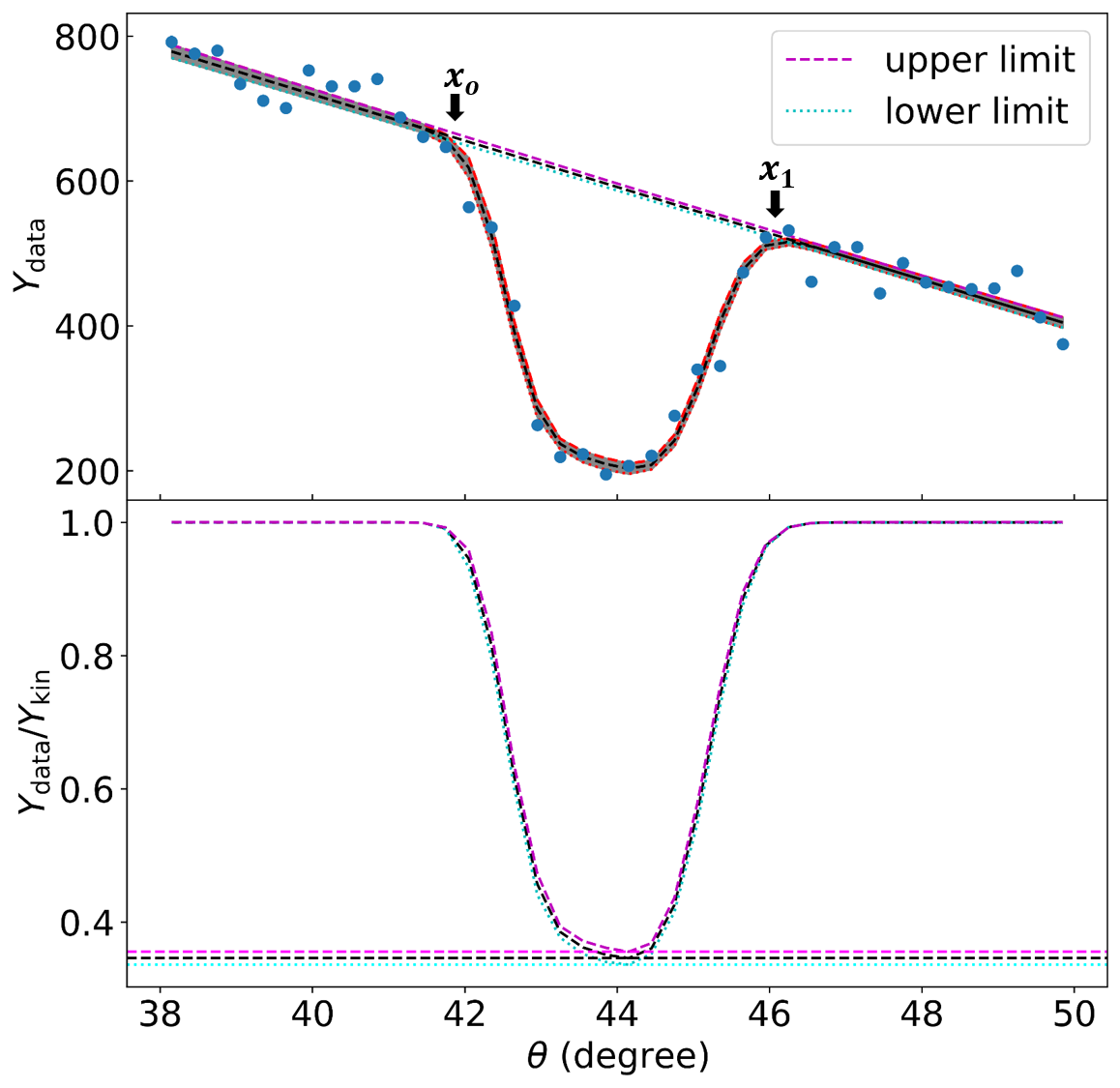}
\caption{Energy independent shadow bar fitting. Top: The total neutron spectra is fitted using Eq.~(\ref{fit_function}) for bar 16 for $^{48}$Ca~+~$^{124}$Sn system at beam energy of 140 MeV/u. Bottom: The fit spectra is scaled by the shape of the spectra around the shadowed regions to determine the neutron background. The dashed and dotted lines are explained in the text.}
\label{eneif}
\end{figure}
\begin{figure}
\includegraphics[width=1.0\columnwidth,angle=0]{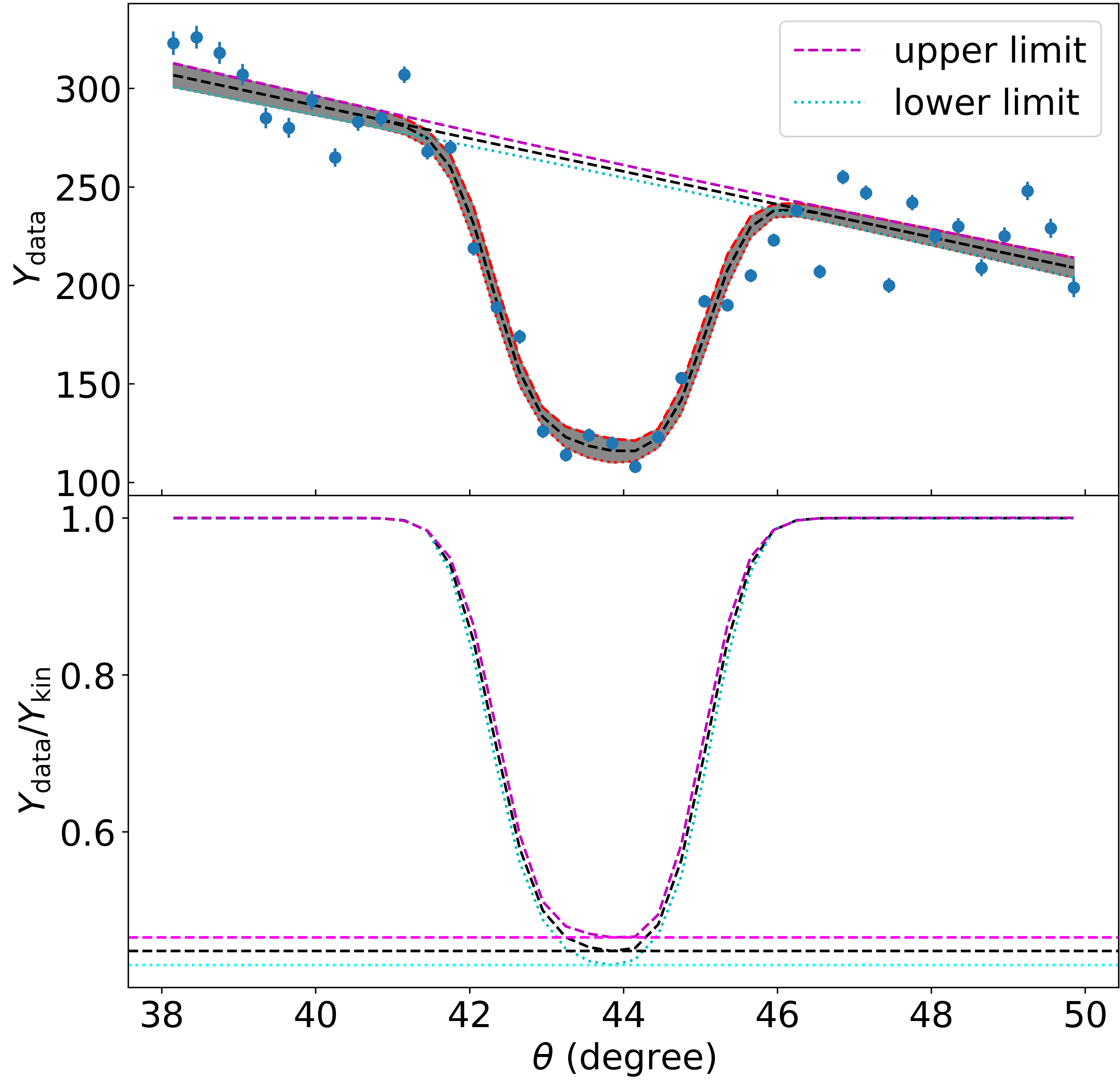}
\caption{Energy dependent shadow bar fitting. The background scattering fraction determined for neutron energies, $E_n$=30~$\pm$~10 MeV for bar 16 for $^{48}$Ca~+~$^{124}$Sn system at beam energy of 140 MeV/u. The lines are as described in Fig.~\ref{eneif}.}
\label{enedf}
\end{figure}

Ideally, the spectrum of a shadowed neutron wall element would include a square gap with a flat bottom, where the bottom of the gap shows the background, and outside the gap is the result of both background and signal. In reality, the edges of the gap are smeared due to the position resolution of the detector as evident from Fig.~\ref{neutron_spectra}. This is represented by Gaussian convoluted square gap characterized by the error function, $\mathrm{erf}$. So, we assume the fit function is characterized by
\begin{equation}\label{fit_function}
\begin{split}
    F(x)=\mathcal{A}*\left(1.0+\mathrm{erf}\left(-\frac{x-x_0}{\sqrt{2}\sigma_0}\right)+\mathrm{erf}\left(-\frac{x-x_1}{\sqrt{2}\sigma_1}\right)\right)\\+(b+c*x),
\end{split}
\end{equation}

where $\mathcal{A}*\left(\mathrm{erf}\left(-\frac{x-x_0}{\sigma_0}\right)+\mathrm{erf}\left(-\frac{x-x_1}{\sigma_1}\right)\right)$ is the component representing the area between $x_0$ and $x_1$~\cite{Coupland}. $x_0$ and $x_1$ are the position of edges of the shadow bar as shown in Fig.~\ref{eneif}, with associated standard deviation $\sigma_0$ and $\sigma_1$ coming from the LANA position resolution. The $\sigma_0$ and $\sigma_1$ values are varied between 0.31 and 0.45 corresponding to the position resolution for bars 8 and 16 in LANA as reported in Ref.~\cite{ZhuKuan}. The $\mathcal{A}+b+c*x$ component, referred as ``kin'' hereafter, represents the expected shape of the neutron spectrum in the absence of the shadow bar~\cite{Coupland}. The neutron spectra for a completely shadowed bar is fitted using Eq.~(\ref{fit_function}), where $\chi^2$ is minimized. The result of the fit for bar 16 for $^{48}$Ca~+~$^{124}$Sn system at beam energy of 140 MeV/u is shown in Fig.~\ref{eneif}. The best fit line is represented by the black dashed line in the figure. To get the background fraction Eq.~(\ref{fit_function}) is scaled by $\mathcal{A}+b+c*x$ as shown in bottom panel of Fig.~\ref{eneif}. The background fraction is then determined at the center between the $x_{0}$ and $x_{1}$ points. The uncertainty in the fitting parameters is propagated to determine the upper limit and lower limits of the best fit line represented by the red dashed and red dotted lines, respectively, in upper panel of Fig.~\ref{eneif}, and their corresponding ``kin'' components represented by dashed pink and dotted cyan lines, respectively. The error bar in neutron background is then determined as the difference between maximum and minimum values inferred as shown in bottom panel of Fig.~\ref{enedf}. This method is referred as ``Method 1" hereafter in the text. 

\par
The neutron background is dependent on neutron energy or equivalently the time of flight. The background fraction in the past has been determined without taking into consideration the energy dependence of neutrons as shown in Fig.~\ref{eneif}, this is equivalent to method 1. In this work, we extend method 1 by dividing the neutron spectra is divided into equal energy bins and  determining the background fraction for each bin. For example, Fig.~\ref{enedf} shows the background scattering fraction determined for neutrons with energies $E_n$=30~$\pm$~10 MeV. The energy bin width of 20 MeV was chosen such that we have sufficient statistics for the fitting process. The figures shown throughout this paper is for $^{48}$Ca~+~$^{124}$Sn system at beam energy of 140 MeV/u for bar 16 unless otherwise stated.

\section{Systematics}
The error in background fraction determined from the algorithm discussed in Sec.~\ref{algorithm} is the statistical error. The dip in the neutron shadowed region might not be characterized by the flat bottom as shown in Fig.~\ref{enedf}. So, another way of calculating the errors in the background fraction is to calculate the standard deviation of the background fraction within the dip region. The error-bars obtained from this approach is much smaller than the statistical error-bars except in the case where we are limited by the statistics.\par

\subsection{Model dependent systematic uncertainty}
In another approach to determine the background fraction, one could envision scaling the neutron spectra with shadow bars by the neutron spectra without shadow bars. Fitting the region around the gap with kin produces the expected shape of the spectrum in the absence of the shadow bar. The spectrum is then scaled by this expected shape of the spectrum in absence of shadow bar. The scaled spectrum is then fitted using the fitting function given by,
\begin{equation}\label{fit_function_coupland}
\begin{split}
    F(x)=S*\left(1.0+\mathrm{erf}\left(-\frac{x-x_0}{\sqrt{2}\sigma_0}\right)+\mathrm{erf}\left(-\frac{x-x_1}{\sqrt{2}\sigma_1}\right)\right)\\+Bg,
    \end{split}
\end{equation}
where $S$ is the signal fraction and $Bg$ is the background fraction. This model assumes a constant neutron background. The uncertainty in the background fraction is the statistical uncertainty in $Bg$. 

The background fractions determined using the above described two methods produces identical results, which is not surprising as both of the methods are based on the assumption that the angular distribution of neutrons is explained by a straight line for a small range of angles. However, the validity of this assumption needs to be verified. Since the two methods discussed so far, we used only the data points from Method 1 to characterize the energy dependent neutron background fraction.

\par
\begin{figure*}
\begin{floatrow}
\ffigbox[\FBwidth]{
       \includegraphics[scale=0.4]{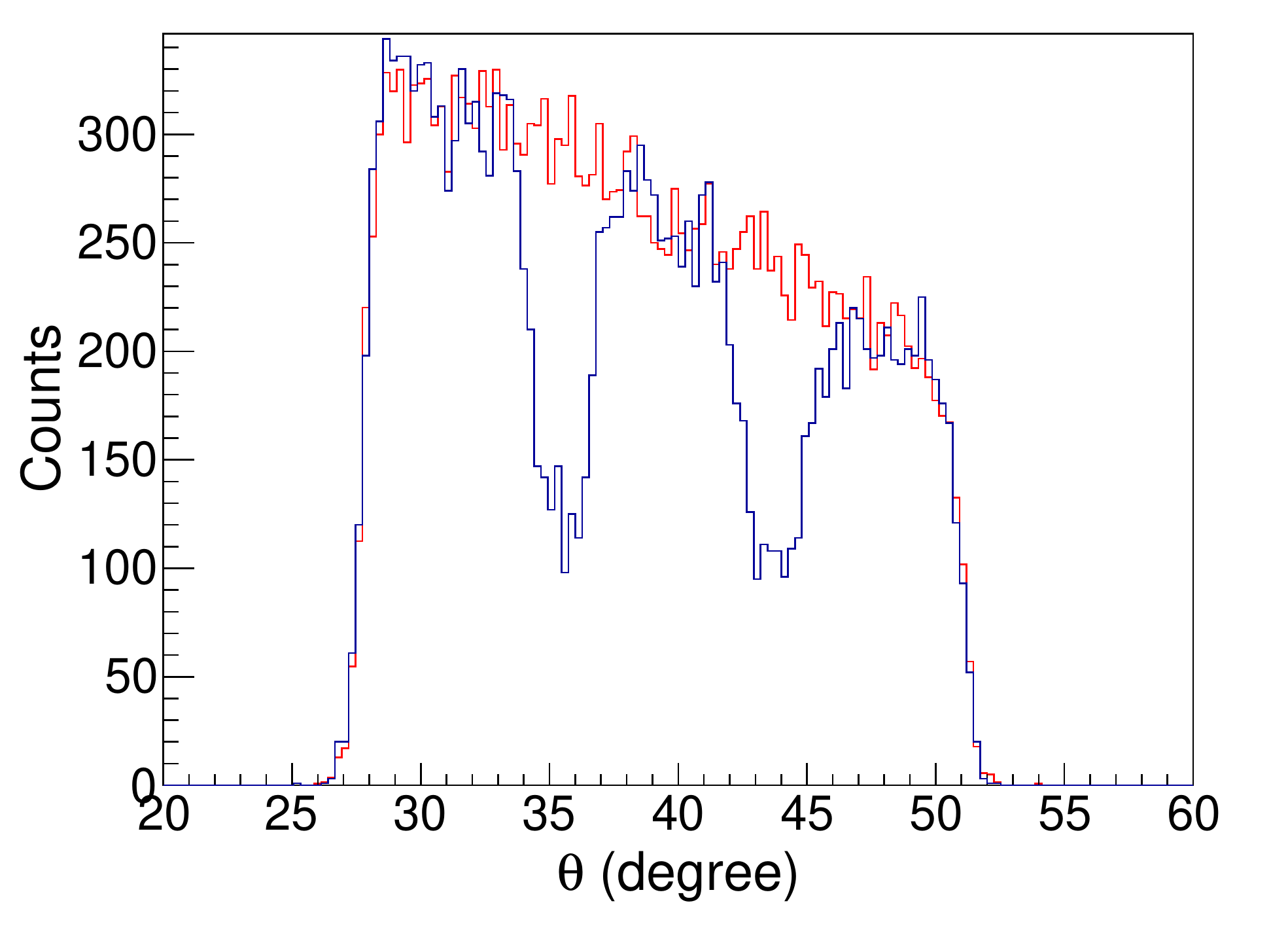} 
 
}{
       \caption{The neutron spectrum, for $E_{n}$=30~$\pm$~10 MeV for bar 16, with and without shadow bars are represented by blue and red histograms respectively. The red histogram is scaled by a factor such that the counts in $37\leq \theta \leq 42$ are same for both histograms. }
        \label{neutron_spectra}
}
\ffigbox[\FBwidth]{
        \includegraphics[scale=0.4]{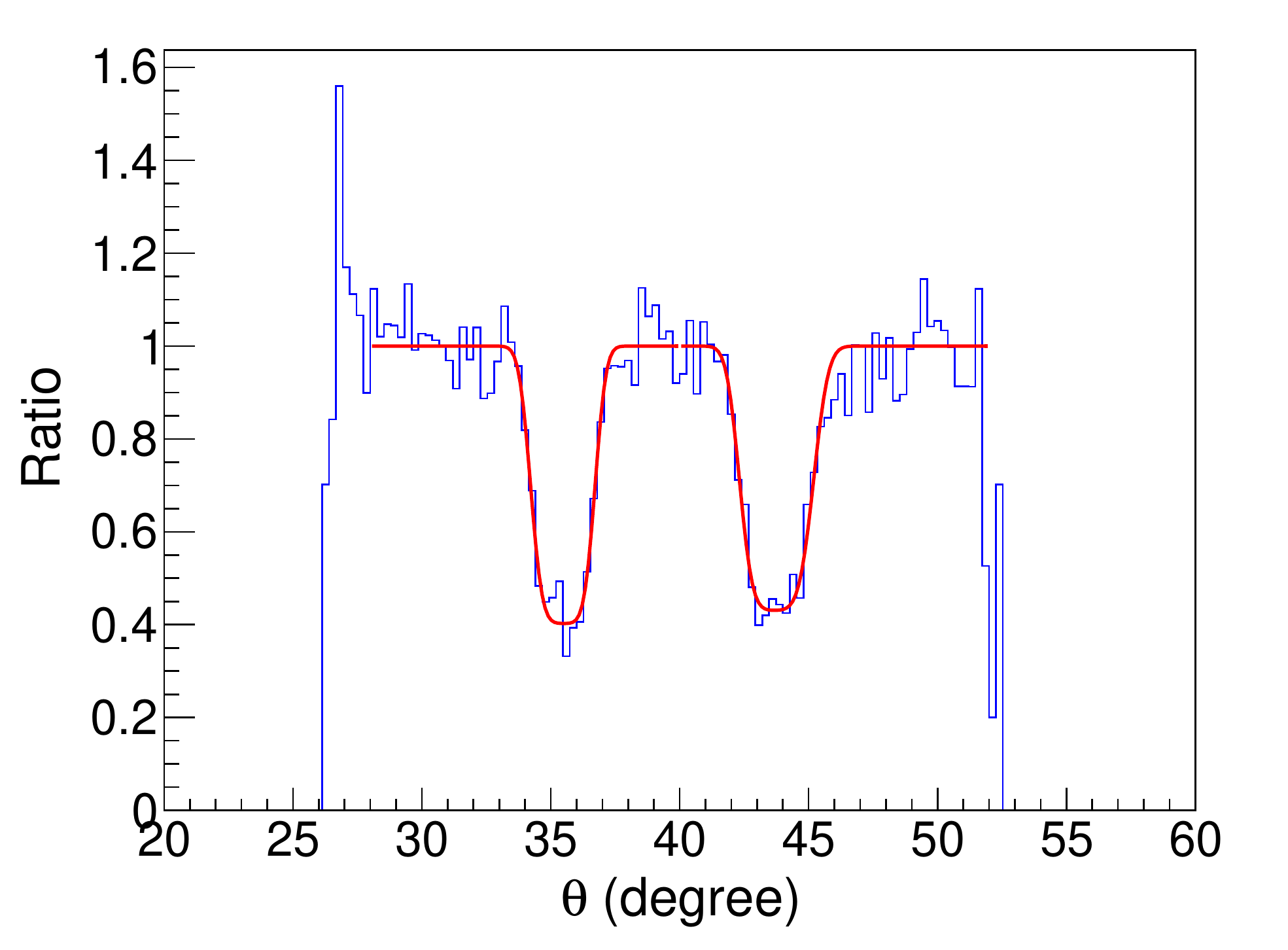} 
        }{
        \caption{The ratio of the neutron spectra, for $E_{n}$=30~$\pm$~10 MeV for bar 16, with and without shadow bars for $^{48}$Ca~+~$^{124}$Sn system at beam energy of 140 MeV/u fitted with the functional form characterized by Eq.~(\ref{fit_function_coupland})}
        \label{fit_spectra}
}
\end{floatrow}
\end{figure*}
\begin{figure}[htp]
\includegraphics[width=1.0\columnwidth,angle=0]{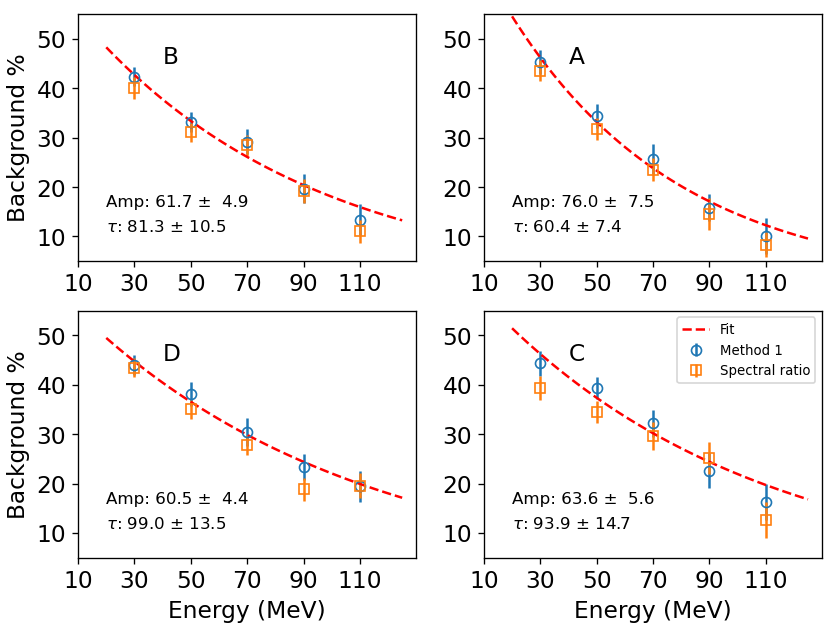}
\caption{Positional dependence of neutron background. The neutron background determined using both Method 1 and spectra ratio method, represented by circles and squares respectively, for $^{48}$Ca~+~$^{124}$Sn system at beam energy of 140 MeV/u. A, B, C, D represents the position of shadow bars. The upper row is for bar 16 and bottom row is for bar 8, respectively. The parameters shown in each figure are obtained from individual fits (dashed red line) of neutron background to data from Method 1 at each position respectively. }
\label{model_comparisons}
\end{figure}

 Instead of scaling the neutron spectra around the shadowed regions with a straight line, we could use the neutron spectra from the experiment where the shadow bars are not used. This is termed as the $``$spectral ratio" method hereafter, where the ratio of the spectra with and without shadow bar is taken and the scaled spectra is fitted with the fitting function characterized by Eq.~(\ref{fit_function_coupland}) to get the background fraction, as shown in Fig.~\ref{neutron_spectra} and Fig.~\ref{fit_spectra}. Figure~\ref{neutron_spectra} shows the observed neutron spectra with and without the shadow bars, represented by the blue and red histograms respectively. The blue histogram is then divided by the red histogram to get the scaled neutron spectra, represented by the histogram in Fig.~\ref{fit_spectra}, which is then fitted using Eq.~(\ref{fit_function_coupland}) to determine the background fraction. The background fractions extracted from this method are shown in Fig.~\ref{model_comparisons} for the shadowed positions along with the background fractions determined from Method 1 for comparison. The background fractions determined by the spectra ratio method are systematically lower than the ones determined using Method 1. This highlights the fact that understanding the angular distribution of neutrons is important to determine the background fraction. However, running experiments with and without shadow bars might not be practically feasible. In the experiment discussed in Section.~\ref{experiment}, there is not enough data to follow the spectral ratio method for all the systems. 

\begin{figure}[htp]
\includegraphics[width=1.0\columnwidth,angle=0]{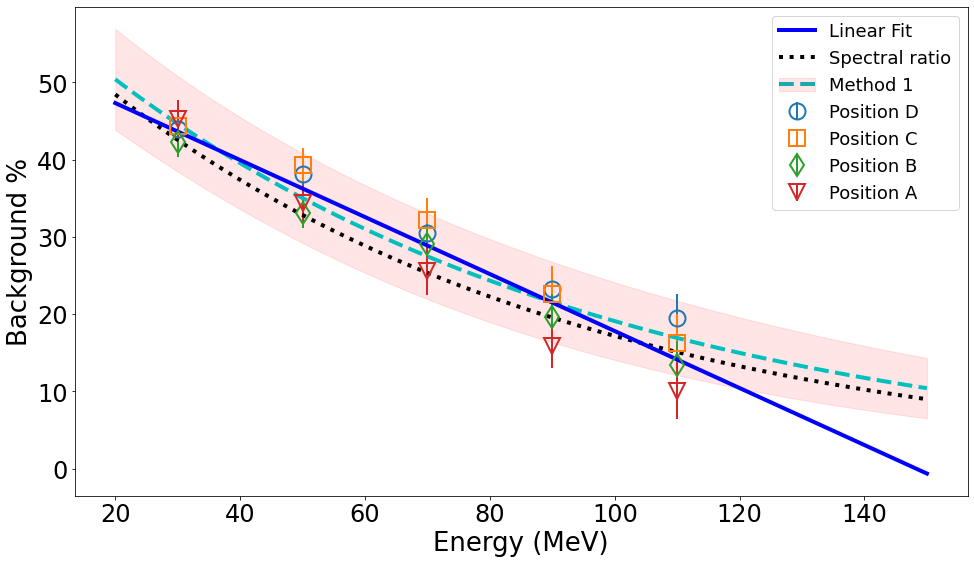}
\caption{Combined fit to neutron background data using the exponential decay function. The shaded region shows the 3-$\sigma$ error band. The cyan dashed line represents the best fit line to all data points obtained from method 1 using the exponential decay function. The black dotted line represents the best fit line to all data points obtained from the spectral ratio method using the exponential decay function. The blue solid line represents the fit to all data points obtained from method 1 using the first order polynomial fitting function which yields a reduced $\chi^2$ value of 1.06 compared to a reduced $\chi^2$ value of 1.30 for exponential decay function represented by dashed cyan line.}
\label{combined_fit}
\end{figure}

The discussion so far about the systematic uncertainties on neutron background have been focused on the method of its determination. The additional contribution comes from the position where the background is being measured. The background fraction is position dependent as evident in Fig.~\ref{model_comparisons} which is not surprising due to the fact that the neutron angular distribution is not flat. The background fraction increases with decreasing neutron energy. The energy dependence of neutron background is also verified by the Monte Carlo Simulation. We used the exponential decay function [$F(E)=\text{Amp}*\text{exp}(-E/\tau)$] to characterize this behaviour. The neutron background measured at each position A, B, C and D are fitted with the exponential decay functions are represented in Fig.~\ref{model_comparisons}. In order to characterize the background as a function of energy for a given system, we used the combined fit to the background fractions determined via method 1 shown in Fig.~\ref{model_comparisons} and come up with a single set of parameters (Amp,$\tau$) with relaxed error bars to describe the positional dependence. It was found that the 3-$\sigma$ error band in the combined fit will explain most of the data points as shown in Fig.~\ref{combined_fit}. The relative error in the determined background is calculated. The systematic uncertainties coming from fitting the backward and forward angles separately is negligible compared to the systematic uncertainty due to the choice of models. The fits to all data points from the spectral ratio method using the exponential decay function will yield a systematically lower background fraction functional curve as represented by the black dashed lines in Fig.~\ref{combined_fit} with similar 3-$\sigma$ bands. The systematic uncertainty due to different models is a function of neutron energy which increases with increasing neutron energy. The systematic uncertainty at neutron energies of 125 MeV due to the choice of model is a maximum of 15$\%$. 

\par
The choice of exponential decay function is motivated by the fact that the background fraction should not be negative at large neutron energies. If we choose the first order polynomial for the combined fit, the background fraction at large neutron energies is negative which is not physically possible, but could be interpreted as no neutron background at those higher energies. The first order polynomial gives the best fit for all the systems compared to the exponential decay as shown in Fig.~\ref{combined_fit}. The reduced $\chi^2$ for exponential fitting function is 1.30 compared to 1.06 that of first order polynomial.  However, the systematic uncertainty due to choice in exponential decay function versus the first order polynomial is less than 5$\%$ where the background fraction data points are available and is not a dominant contributor to the overall systematic uncertainty.

Although the systematic uncertainty at large neutron energies is higher, but the fact that the background fraction at these energies is much smaller, the overall uncertainty to the neutron spectra at large energies is less than 6$\%$ for most of the systems we analyzed.
\section{Conclusion}
In summary, we describe the method of determining the background fraction using the shadow bars in the heavy-ion collision experiment. The systematic uncertainty of the method is described and found that understanding the neutron angular distribution is important to determine the background fraction. The position or angle dependence of the neutron background is observed. We would recommend measuring the neutrons with and without the shadow bars to determine the neutron background fraction more accurately with spectral ratio method. The non-linear angular distributions of neutrons impacts the neutrons background determined using method 1.
\section{ACKNOWLEDGEMENTS}
This work was supported in part by the U.S. National Science Foundation under Grants PHY-1565546, PHYS-1712832, and PHY-2110218; in part by the U.S. Department of Energy (the Office of Science) under Grant DE-NA0003908; and in part by the National Research Foundation of Korea (NRF) under Grant 2016K1A3A7A09005578, Grant 2018R1A5A1025563, and Grant 2013M7A1A1075764.

\bibliography{references}
\end{document}